\documentclass[aps,prb,showpacs,superscriptaddress,twocolumn]{revtex4-1}

\usepackage{amsfonts}
\usepackage{subfigure}
\usepackage{amsmath}
\usepackage{txfonts}
\usepackage{amssymb}
\usepackage{amsbsy} %for sigma in bold face
\usepackage{epsfig}
\usepackage{graphicx}
\usepackage{epstopdf}

\usepackage{mathdots} %from Fei Song

\def\be{\begin{equation}} \def\ee{\end{equation}}
\def\bea{\begin{eqnarray}} \def\eea{\end{eqnarray}}

\def\be{{\bf e}}

\begin{document}
\title{Effect of incommensurate potential on nodal-link semimetals}
\author{Yucheng Wang}
\affiliation{Beijing National
Laboratory for Condensed Matter Physics, Institute of Physics,
Chinese Academy of Sciences, Beijing 100190, China}
\affiliation{School of Physical Sciences, University of Chinese Academy of Sciences, Beijing, 100049, China}
\author{Haiping Hu}
\affiliation{Department of Physics, The University of Texas at Dallas, Richardson, Texas 75080, USA}
\author{Shu Chen}
\email{schen@iphy.ac.cn}
\affiliation{Beijing National
Laboratory for Condensed Matter Physics, Institute of Physics,
Chinese Academy of Sciences, Beijing 100190, China}
\affiliation{School of Physical Sciences, University of Chinese Academy of Sciences, Beijing, 100049, China}
\affiliation{Collaborative Innovation Center of Quantum Matter, Beijing, China}
%\date{\today}

\begin{abstract}
We explore the effect of an incommensurate potential on nodal-link semimetals. For system with surface states, we show the nodal-link semimetal undergoes a series of phase transitions: first into a metallic phase, then to a loop semimetal phase and back to a metallic phase with increasing the strength of incommensurate potential. The phase transitions are unveiled by analyzing the properties of energy spectra and wave functions. We also study the system without surface states and find that the Fermi surface evolves from lines to loops before the system enters into a metallic phase when increasing the incommensurate potential strength.
\end{abstract}
\pacs{64.60.-i, 71.30.+h, 72.15.Rn, 71.55.Ak}
%64.60.-i:general studies of phase transitions; 71.30.+h:metal-insulator transitions and other electronic transitions;
%72.15.Rn:localization effects(Anderson or weak localization);71.55.Ak:metal,semimetals, and alloys
\maketitle
%%%%%%%%%%%%%%%%%%%%%%%%%
\section{Introduction}
%%%%%%%%%%%%%%%%%%%%%%%%%%
The recent discoveries of topological phases of matters, which contain both gapped systems and gapless semimetals \cite{Nagaosa,Fang}, have vastly deepened our understanding of quantum phases of matter. The gapped topological states, including topological insulators and topological superconductors \cite{Kane,Zhang}, can be well described by the underlying symmetries \cite{Chiu,class1,class2} and associated topological invariants\cite{Chiu}. While for the gapless semimetals, the crossings of the conduction and valance band may occur at some discrete points in Brillouin zone (BZ), such as Weyl and Dirac semimetal \cite{Murakami,Wan,Burkov,Yang,Xu,Halasz,Young,Huang,Lv,Hasan,Hasan2,Yang2}, or along some closed loops, corresponding to topological nodal line semimetals \cite{Burkov2,Fang2,Chan2016}. There exist some possible configurations for two nodal loops. They can be isolated, touched at one point, or linked with each other, known as nodal link semimetal \cite{Hou,Yan,Chang,Ezawa}. The nontrivial linking of the nodal lines is described by a Hopf link number and may bring exotic topological properties to the system. A model Hamiltonian for Hopf insulators in a solid-state quantum simulator has been implemented and the experimental observation of their topological properties was also reported \cite{DuanLM17}.

The effect of disorder on topological state is an important question, especially for topological semimetals, due to the lack of protection of a finite energy gap and vanishing density of states (DOS) at the Fermi level. For Dirac \cite{Herbut,Sondhi,Sarma1,Sarma2} and Weyl \cite{Huang2,Ominato,Brouwer,Ryu,Xie,Hughes,Roy,Gurarie} semimetals, its have been found that the semimetal phases are robust against weak disorder and generally the system would first enter into a metal phase before becoming an insulator by increasing the strength of disorder. The effect of an incommensurate potential on the Weyl semimetals was firstly studied in Ref. \cite{Wang} by adding a quasiperiodic potential along one direction of the lattice, which unveils a transition from a Weyl semimetal to a metal driven by the incommensurate potential.  For three-dimensional incommensurate potentials, it has been demonstrated recently by Pixley et al.\cite{Pixley} that there exist Weyl semimetal (WSM)-metal-semimetal-metal phase transitions by increasing the incommensurate potential strength. The incommensurate potential is quasiperiodic, lying between a random potential and periodic potential and has been widely used in cold atom experiments \cite{Roati}. It can induce many interesting phenomenon, such as ballistic transport \cite{Sokoloff} and finite localized-delocalized transition in one dimension \cite{Azbel,AA} distinct from the random potential.

Taking advantages of these features, a natural and interesting question is then what is the fate of the nodal-link semimetal by applying such an incommensurate potential? To answer this question, we consider the simplest case with incommensurate potential applied only in one spatial direction. The introduced incommensurate potential only destroys the translation invariance in one dimension, whereas momentums in other two direction are still good quantum numbers. Such a choice not only simplifies  our calculation and overcomes the size difficulty, but also provides important insights \cite{Takane,Wang} towards the disorder effects of nodal-link semimetals under different boundary conditions.

This paper is organized as follows: in Sec. \ref{model}, we introduce the three dimensional Hamiltonian with nodal-link Fermi surface in the presence of a one dimensional incommensurate potential. We then consider two different cases depending on the existence of surface states (corresponding to take open boundary conditions (OBC) in the $z$ direction or $y$ direction of the system) and analysis the phase transitions by increasing the strength of incommensurate potential. A brief summary is given in Sec. \ref{conclusion}.

%%%%%%%%%%%%%%%%%%%%%%%%
\section{Model and results}
\label{model}
\begin{figure}
\includegraphics[height=65mm,width=80mm]{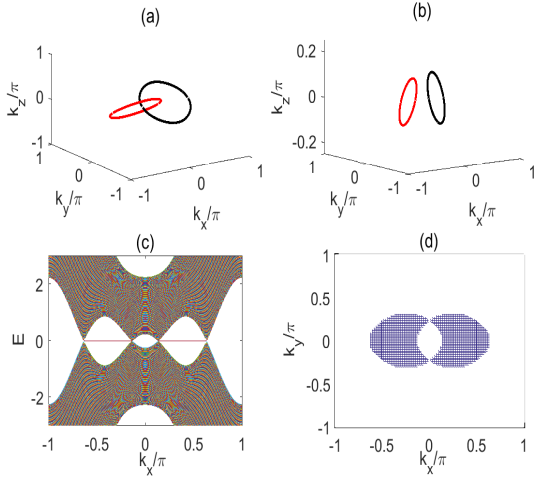}
\caption{\label{01}(Color online)
Nodal rings of our system (Eq. \ref{ham-3}) with $V=0$, (a) $m_0=2.5$ and (b) $m_0=3.2$. (c) the energy spectra as a function of $k_x$ under OBC along $z$ direction for fixed $k_y=0$ in the absence of the incommensurate potential and (d) the corresponding cross section of energy dispersion $E(k_x,k_y)$ at $E=0\pm 0.03$ (with considering the finite size effect). Here $L=300$.}
\end{figure}

%%%%%%%%%%%%%%%%%%%%%%%%%
A general two-band Bloch Hamiltonian can be written as:
\begin{equation}
H_0(\textbf{k})=d_0(\textbf{k})\textbf{1}+d_x(\textbf{k})\tau_x+d_y(\textbf{k})\tau_y+d_z(\textbf{k})\tau_z,
\label{ham-1}
\end{equation}
where $\textbf{k}=(k_x,k_y,k_z)$ and $\tau_i$ ($i=x,y,z$) is Pauli matrix. For a spinless system, the time reversal operator $T$ and spatial inversion operator $P$ act on the Hamiltonian through $TH(\textbf{k})T^{-1}=H^{*}(-\textbf{k})$ and $PH(\textbf{k})P^{-1}=H(-\textbf{k})$, respectively. We assume the $H(\textbf{k})$ satisfy the $PT$ symmetry \cite{Fang,Zhao} $PTH(\textbf{k})(PT)^{-1}=H^{*}(\textbf{k})$, which requires that $d_y(\textbf{k})=0$. We further set $d_0(\textbf{k})=0$ without loss of generality. As a frequently-used example, we take \cite{Moore,Deng1,Deng2,Deng3,Kennedy,Liu,Yan}:
%\begin{equation}
%\begin{aligned}
\bea
d_x &=& 2\sin k_x \sin k_z+2 \sin k_y (m_0-\sum_{i=x,y,z} \cos k_i), \nonumber\\
d_z &=& \sin^2k_x+\sin^2k_y-\sin^2k_z \nonumber\\
& &  -(m_0-\sum_{i=x,y,z} \cos k_i)^2.
\label{ham-2}
\eea
%\end{aligned}
%\end{equation}
We consider the half-filling case and the zero energy Fermi surface is determined by $d_x=d_z=0$. For $1<m_0<3$, the Fermi surface contains two linked rings as shown in Fig.~\ref{01}(a) with $m_0=2.5$, whereas it contains two unlinked rings when $m_0>3$ as shown in Fig.~\ref{01}(b) with $m_0=3.2$.   One of the rings satisfies $k_y=k_z$ and $\sin k_x=m_0-\sum_{i=x,y,z} \cos k_i$, the other satisfies $k_y=-k_z$ and $\sin k_x=\sum_{i=x,y,z} \cos k_i-m_0$.  If we take periodic boundary conditions (PBC) along the $x$ and $y$ directions and OBC in the $z$ direction, $k_z$ will be no longer a good quantum number. In Fig.~\ref{01}(c), we show the energy spectra as a function of $k_x$ by fixing $k_y=0$, where the $L_x\times L_y\times L_z$ cubic lattice is set as $L_x=L_y=100a$ unless otherwise stated and $L_z=La$. For convenience, we set the lattice constant $a=1$ and $m_0=2.5$ in the following context. We can clearly observe the emergence of surface states connecting the projected nodal points. The corresponding zero energy surface flat bands localized at $z=0$ and $z=L_z=La$ are shown in the blue area of Fig.~\ref{01}(d).

%%%%%%%%%%%%%%%%%%%%%%%%%%%%%%%%%%%%%%%%%%%%%%%
%%%%%%%%%%%%%%%%%%%%%%%%%%%%%%%%%%%%%%%%%%%%%%%%
%%%%%%%%%%%%%%%%%%%%%%%%%%%%%%%%%%%%%%%%%%%%%%%
\begin{figure}
\includegraphics[height=55mm,width=65mm]{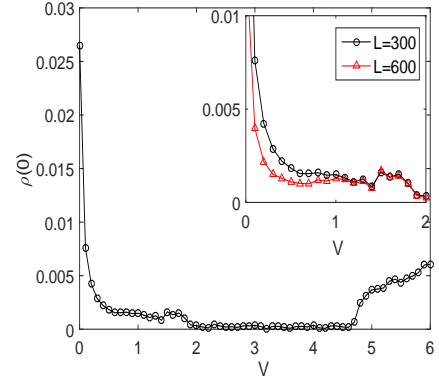}
\caption{\label{01s}(Color online)
DOS $\rho(0)$ versus the strength of incommensurate potential $V$ for the system with $L=300$ and the inset is that $\rho(0)$ as a function of $V$ for $L=300$ and $L=600$.}
\end{figure}
%%%%%%%%%%%%%%%%%%%%%%%%%%%%%%%%%%%%%%%%%%%%%%%%

We then add an incommensurate potential along the $z$ direction. The Hamiltonian becomes
\begin{equation}
\begin{aligned}
H=\sum_{k_x,k_y,k_z}\mathbf{a}^{\dagger}H_0\mathbf{a} + V \sum_{k_x,k_y,l} \cos(2\pi\alpha l)n_{k_x,k_y,l},
\label{ham-3}
\end{aligned}
\end{equation}
where $\mathbf{a}=[a^{A}_{k_x,k_y,k_z}, a^{B}_{k_x,k_y,k_z}]^{T}$, $n_{k_x,k_y,l}=(a^{A\dagger}_{k_x,k_y,l}a^{A}_{k_x,k_y,l}+a^{B\dagger}_{k_x,k_y,l}a^{B}_{k_x,k_y,l})$ and $\alpha$ is an irrational number, which is chosen as $\alpha=(\sqrt{5}-1)/2$ without loss of generality. For fixed $k_x$ and $k_y$, the Hamiltonian (\ref{ham-3}) can be written in real space by using Fourier transformation: $a^{\gamma}_{k_x,k_y,l}=\frac{1}{\sqrt{L_z}}\sum_{k_z}e^{i(k_zla)}a^{\gamma}_{k_x,k_y,k_z}$, where $\gamma=A, B$ label the unequivalent sublattice of the $2\times 2$ matrix and $a^{A}_{k_x,k_y,l}$ ($a^{B}_{k_x,k_y,l}$) is the annihilation operator on the sublattice A (B) of the $l-th$ layer along the $z$ direction. The $n$-th eigenstate of the Hamiltonian (\ref{ham-3}) can be generally represented as $|\Psi_n\rangle=\sum_{l}(\psi_{n,l,A}a^{A\dagger}_{k_x,k_y,l}+\psi_{n,l,B}a^{B\dagger}_{k_x,k_y,l})|0\rangle$. By using the Schr\"{o}dinger equation $H|\Psi\rangle=E_{n,k_x,k_y}|\Psi\rangle$, we can obtain the eigenvalues of this system (see the Appendix A).
%%%%%%%%%%%%%%%%%%%%%%%%%%%%%%%%%%%%%%%%%%%%%%%
\begin{figure*}
\includegraphics[height=80mm,width=160mm]{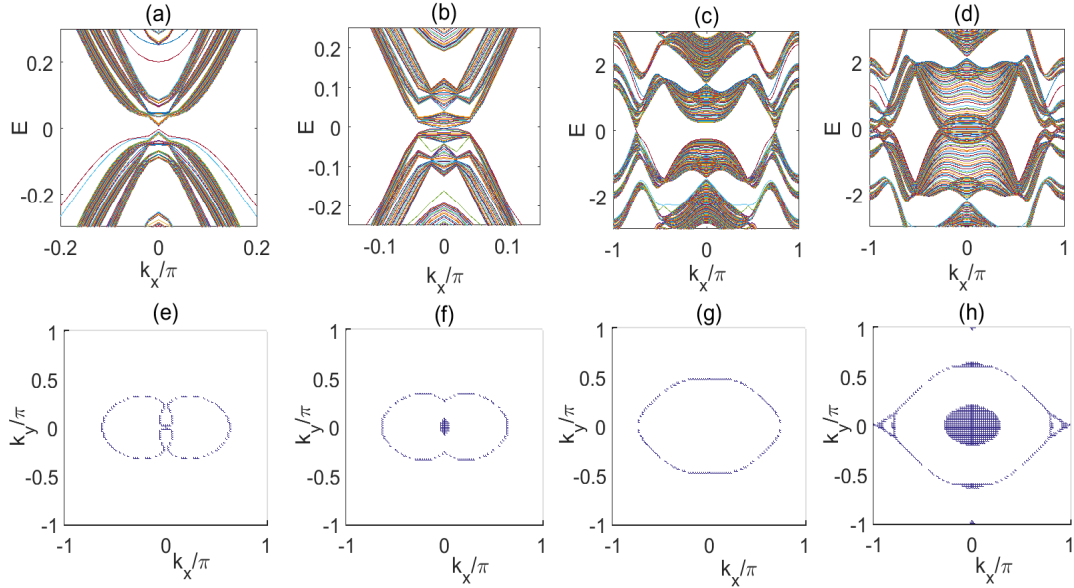}
\caption{\label{02}(Color online)
 The energy spectra as a function of $k_x$ under OBC along $z$ direction for fixed $k_y=0$ with the incommensurate potential strength (a) $V=1$, (b) $V=1.5$, (c) $V=4$ and (d) $V=6$. The corresponding cross section of energy dispersion $E(k_x,k_y)$ at $E=0\pm 0.03$ (with considering the finite size effect) for this system with (e) $V=1$, (f) $V=1.5$, (g) $V=4$ and (h) $V=6$. Here $L=300$. For (a) and (b), to make the details clearer and increase the resolution, we fix the range of $k_x$ as $(-0.2\pi,0.2\pi)$ and $L_x=1000$.}
\end{figure*}
%%%%%%%%%%%%%%%%%%%%%%%%%%%%%%%%%%%%%%%%%%%%%%%%
%%%%%%%%%%%%%%%%%%%%%%%%%%%%%%%%%%%%%%%%%%%%%%%
\begin{figure}
\includegraphics[height=40mm,width=85mm]{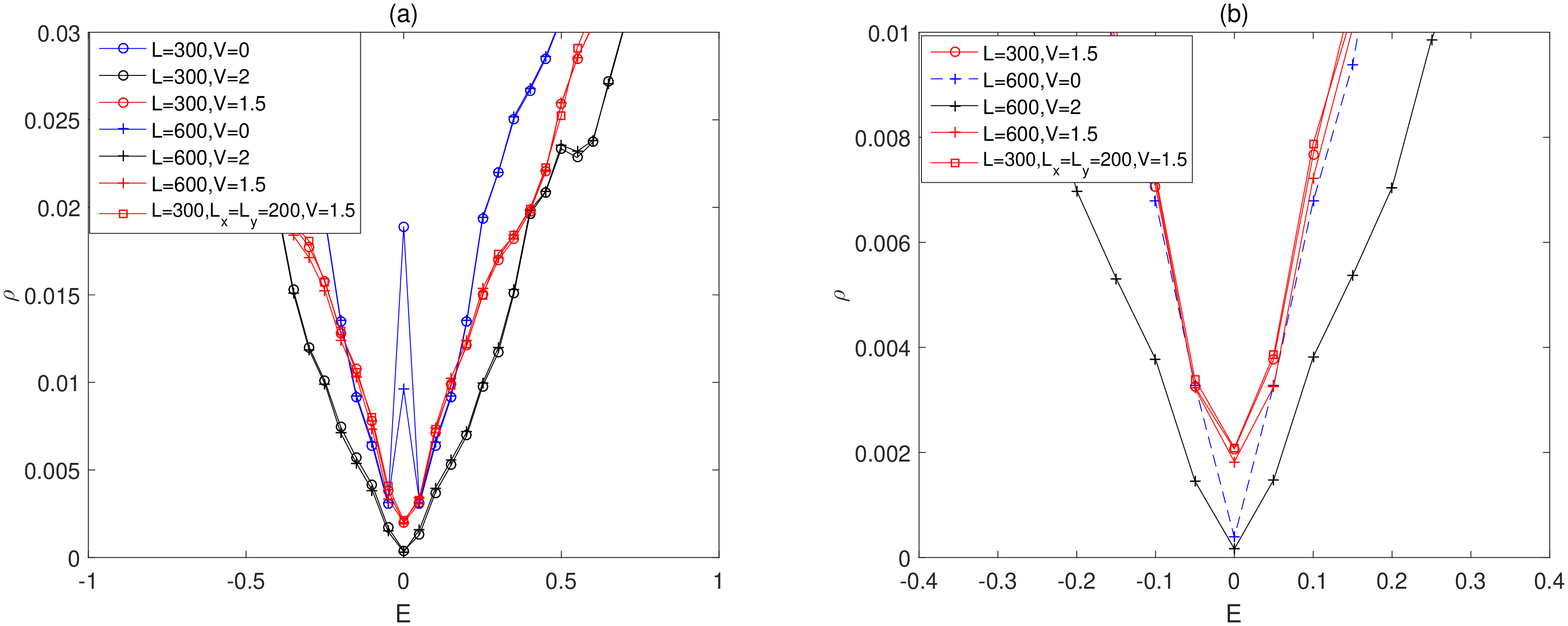}\\
\includegraphics[height=40mm,width=85mm]{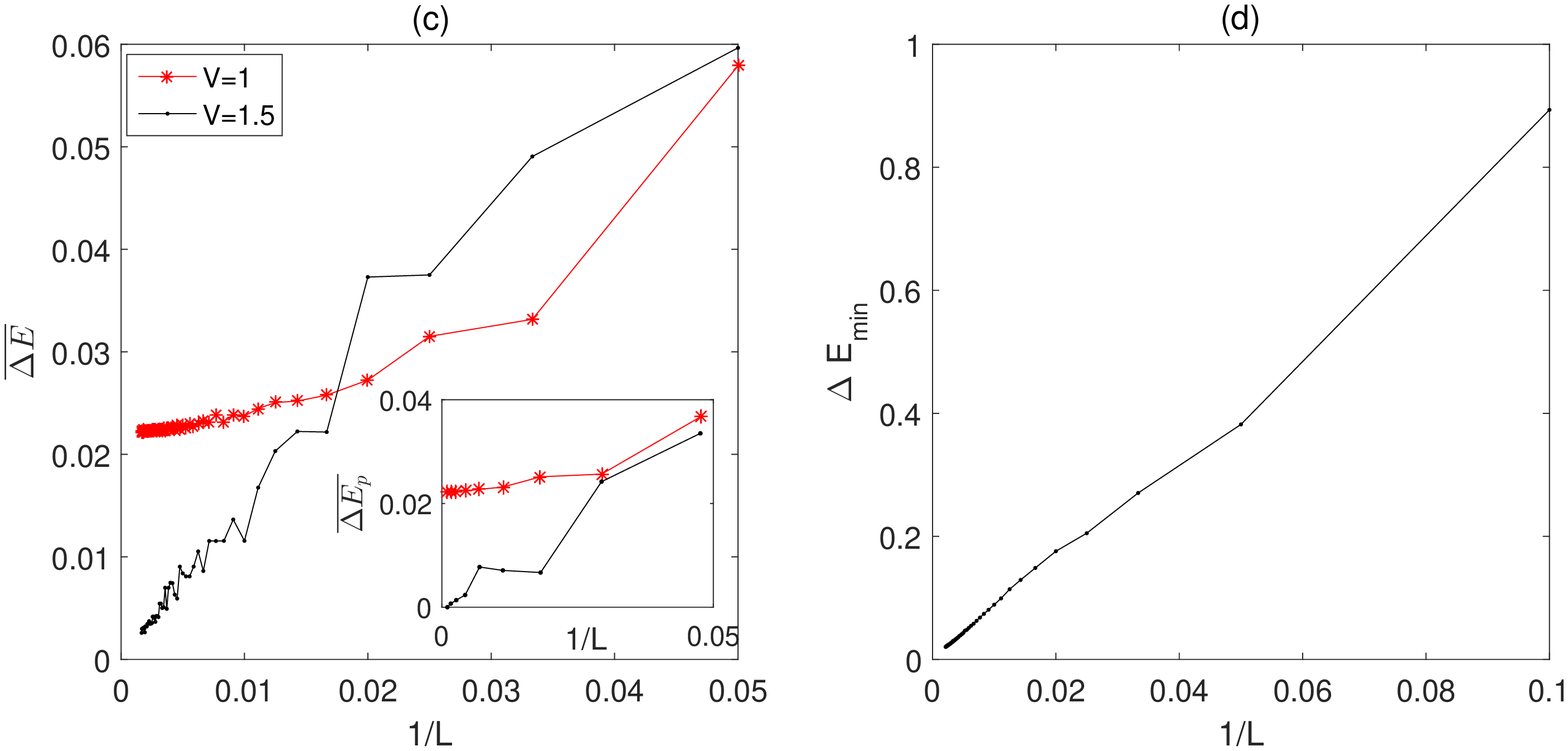}
\caption{\label{03}(Color online)
  DOS as a function of energy $E$ for various values of incommensurate potential strength $V$ and system size under (a) OBC and (b) PBC in the $z$ direction.
  (c) $\overline{\Delta E}$ versus $1/L$ for $V=1$ and $V=1.5$ respectively. The inset shows the corresponding $\overline{\Delta E_p}$ under PBC. (d) $\Delta E_{min}$ as a function of $1/L$ for fixed $V=4$.}
\end{figure}
%%%%%%%%%%%%%%%%%%%%%%%%%%%%%%%%%%%%%%%%%%%%%%%%

To see how the nodal-link semimetals are changed with the increasing of incommensurate potential strength, we introduce the DOS defined as
\begin{equation}
\rho(E) = \frac{1}{2L_xL_yL_z}\sum_{l=1}^{2L_z}\sum_{i_x=-\frac{L_x}{2}}^{\frac{L_x}{2}-1}\sum_{i_y=-\frac{L_y}{2}}^{\frac{L_y}{2}-1}\delta(E-E_{l,i_x,i_y}).
\label{dos}
\end{equation}
where $k_x=\frac{2\pi}{L_x}i_x$ and $k_y=\frac{2\pi}{L_y}i_y$ are the quantized momenta along periodic directions. $E_{l,k_x,k_y}$ is the $l$-th eigenvalue of the Hamiltonian at $(k_x,k_y)$. Numerically, the $\delta$ function in Eq. (\ref{dos}) can be approximated by using a Gaussian function\cite{Fehske} $\frac{1}{\sqrt{\pi\sigma^2}}\exp(-\frac{x^2}{\sigma^2})$ with a small $\sigma$, which is set as $0.01$ in the subsequent calculations.

The zero energy DOS $\rho(0)$ with respect to the strength of incommensurate potential $V$ is shown in Fig.~\ref{01s}. Without the incommensurate potential, $V=0$, the zero energy DOS takes a finite value, consistent with the existence of middle surface states. When the incommensurate potential is added, these surface states bend downward to $E<0$, as shown in Fig.~\ref{02}(a), where we show the energy spectra as a function of $k_x$ for fixed $k_y=0$ and $V=1$. In other words, the flat surface states will become non-flat drumhead surface states when adding a small incommensurate potential.(Further increasing the strength of this incommensurate potential, the surface states will be absorbed into the bulk, see the Appendix B for detailed discussions.) The $E=0$ Fermi surface shrinks to lines on the surface BZ as shown in Fig.~\ref{02}(e) and $\rho(0)$ drops sharply as shown in Fig.~\ref{01s}. By further increasing $V$, for instance, $V=1.5$, as shown in Fig.~\ref{02}(b), one can find that the touched segment of the upper and lower bands near $k_x=k_y=0$ is no longer point. Fig.~\ref{02}(f) corresponds to the cross section of energy dispersion $E(k_x,k_y)$ at $E=0\pm 0.03$ (with considering the size effect) with $V=1.5$ and we see that the two-band touched segment is no longer composed of lines and there exist a surface for $E=0$. As a result, $\rho(0)$ is a finite value in the thermodynamical limit for this case. This can also be revealed in Fig.~\ref{01s}, where no obvious difference happen for the zero-energy DOS $\rho(0)$ with different system length when $1.5<V<1.8$.

If the incommensurate potential strength $V$ continues to increase to $V>1.8$, though the cross segment near $k_x=0$ would open an energy gap, there emerges two gap closing point at finite $k_x$ as shown in Fig.~\ref{02}(c) with $V=4$. The $\rho(0)$ will drop to zero again as shown in Fig.~\ref{01s} and the cross segment of the upper and lower bands now becomes a loop  in the two dimensional (2D) BZ as shown in Fig.~\ref{02}(g), i.e., this is a loop semimetal phase. Further increasing  the strength of the incommensurate potential to $V >4.6$, the upper and lower bands will touch again as shown in Fig.~\ref{02}(d) with $V=6$. The touched segment becomes a 2D region in the BZ as shown in Fig.~\ref{02}(h). The corresponding $\rho(0)$ becomes a finite value as shown in Fig.~\ref{01s} and this system enters into the metallic phase again.

To validate the above discussions, we show the DOS as a function of energy for various values of incommensurate potential strength $V$ and system size in Fig.~\ref{03}(a). For $V=0$, we see that $\rho(0)$ is bigger than the values of $\rho(0)$ for other $V$s because of the existence of flat surface states, which will become small when increasing $L$. For $V=1.5$, we see that $\rho(0)$ is a finite non-zero value and remains almost unchanged with changing $L$ or $L_x, L_y$. For $V=2$, $\rho(0)$ approximates to zero because this system becomes a semimetal and doesn't exist the flat surface states. To remove the effect of surface states, we show the corresponding DOS under PBC in Fig.~\ref{03}(b). One can see that the $\rho(0)$ for $V=0$ is close to zero because it is a nodal-link semimetal. From these values of $\rho(0)$ in $V=0$, $V=1.5$ and $V=2$, we can see that $\rho(0)$ changes from $0$ to a finite non-zero value and back to zero, which corresponds to that this system changes from the semimetal to a metallic phase and back to a semimetallic phase.

We further analyze the corresponding energy gaps. For fixed $k_x$ and $k_y$, we can define the energy gap of the upper band and the lower band near $E=0$ as
$\Delta E_{k_x,k_y}=E_{L+2,k_x,k_y}-E_{L,k_x,k_y}$, where $E_{L,k_x,k_y}$ and $E_{L+2,k_x,k_y}$ are the $L$-th and $(L+2)$-th eigenvalues at ($k_x$,$k_y$) and the eigenvalues are in ascending order. Note that we use the $(L+2)$-th  instead of $(L+1)$-th state because at some specific $(k_x,k_y)$ the latter may correspond to the surface state, which bend downwards to $E <0$ by increasing the incommensurate potential strength. We take a small region near $k_x=0, k_y=0$, e.g., $|k_x|\leq 0.01\pi,|k_y|\leq 0.01\pi$ and divide this region into $N\times N$ equal parts. Then the mean energy gap is
\begin{eqnarray}
\overline{\Delta E}=\frac{1}{N^2}\sum_{k_x,k_y}\Delta E_{k_x,k_y}.
\end{eqnarray}
with the summation over all the $k_x$ and $k_y$ in the chosen small region. If $\overline{\Delta E}$ approaches zero in the thermodynamical limit, the size of the zero energy Fermi surface is then bigger than this region, while if $\overline{\Delta E}$ approaches a finite value in the thermodynamical limit, the size of the Fermi surface is smaller than the chosen region.
We divide the region $|k_x|\leq 0.01\pi$ and $|k_y|\leq 0.01\pi$ into $40\times 40$ equal parts and show $\overline{\Delta E}$ as a function of $1/L$ in Fig.~\ref{03}(c).
It is obvious the mean energy gap  $\overline{\Delta E}$ tends to zero when $L\rightarrow \infty$ for $V=1.5$, which means the size of the fermi surface is bigger than $0.02\pi\times 0.02\pi$ and this system enters into a metallic phase. As a comparison,  we also show the mean energy gap $\overline{\Delta E}$ for $V=1$. It tends to a finite value when $L\rightarrow \infty$. We can further narrow this region almost to zero. In the whole process, $\overline{\Delta E}$ would not tend to zero when $L\rightarrow \infty$. However, $\Delta E$ tends to zero in the thermodynamical limit when only considering the states at $k_x=0, k_y=0$ for $V=1$, indicating that the system is a semimetal phase. To eliminate the effect of the boundary states, we consider the corresponding mean energy gap $\overline{\Delta E_p}$ under PBC, here $\Delta E_{p}=E_{L+1}-E_{L}$ for fixed $k_x$ and $k_y$. The irrational number $\alpha=\frac{\sqrt{5}-1}{2}$ can be approached by the Fibonacci numbers via the relation
 $\lim_{n \rightarrow \infty}\frac{F_{n-1}}{F_{n}}=\alpha$,
where $F_{n}$ is defined recursively by  $F_{n+1}=F_{n-1}+F_{n}$, with $F_0=F_1=1$ \cite{Kohmoto}.
The system size $L$ is chosen as $F_n$. In the inset of Fig.~\ref{03}(c), we show the $\overline{\Delta E_p}$ versus $1/L$ for $V=1$ and $V=1.5$ respectively, where the region $|k_x|\leq 0.01\pi$ and $|k_y|\leq 0.01\pi$ is still divided into $40\times 40$ equal parts. It can be seen that $\overline{\Delta E_p}$ tends to zero for $V=1.5$ and it tends to a finite value for $V=1$ when $L\rightarrow \infty$, where the finite value is same with the value $\overline{\Delta E}$ obtained by using OBC.
For $V\approx 1.8\sim 4.6$ case, the zero-energy DOS $\rho(0)$ tends to zero. In Fig.~\ref{03}(d), we plot the minimum gap $\Delta E_{min}=\text{Min}\{\Delta E_{k_x,k_y}\}$ with respect to $1/L$ for fixed $V=4$. It tends to zero in thermodynamical limit, indicating that the energy gap would not open along the nodal loop. The system is in a loop semimetal phase.

%%%%%%%%%%%%%%%%%%%%%%%%%%%%%%%%%%%%%%%%%%%%%%%
\begin{figure}
\includegraphics[height=80mm,width=80mm]{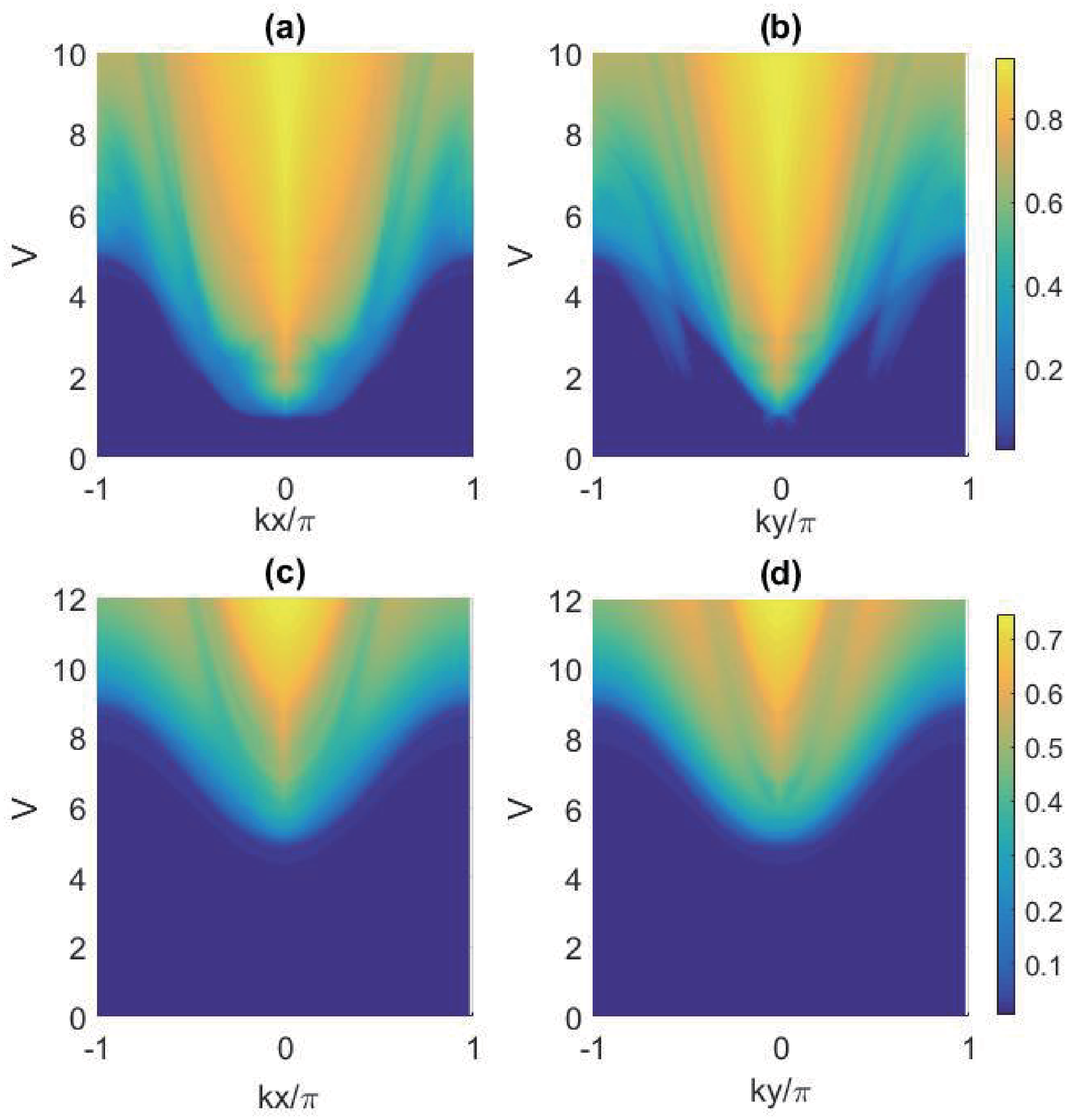}\\
\includegraphics[height=40mm,width=80mm]{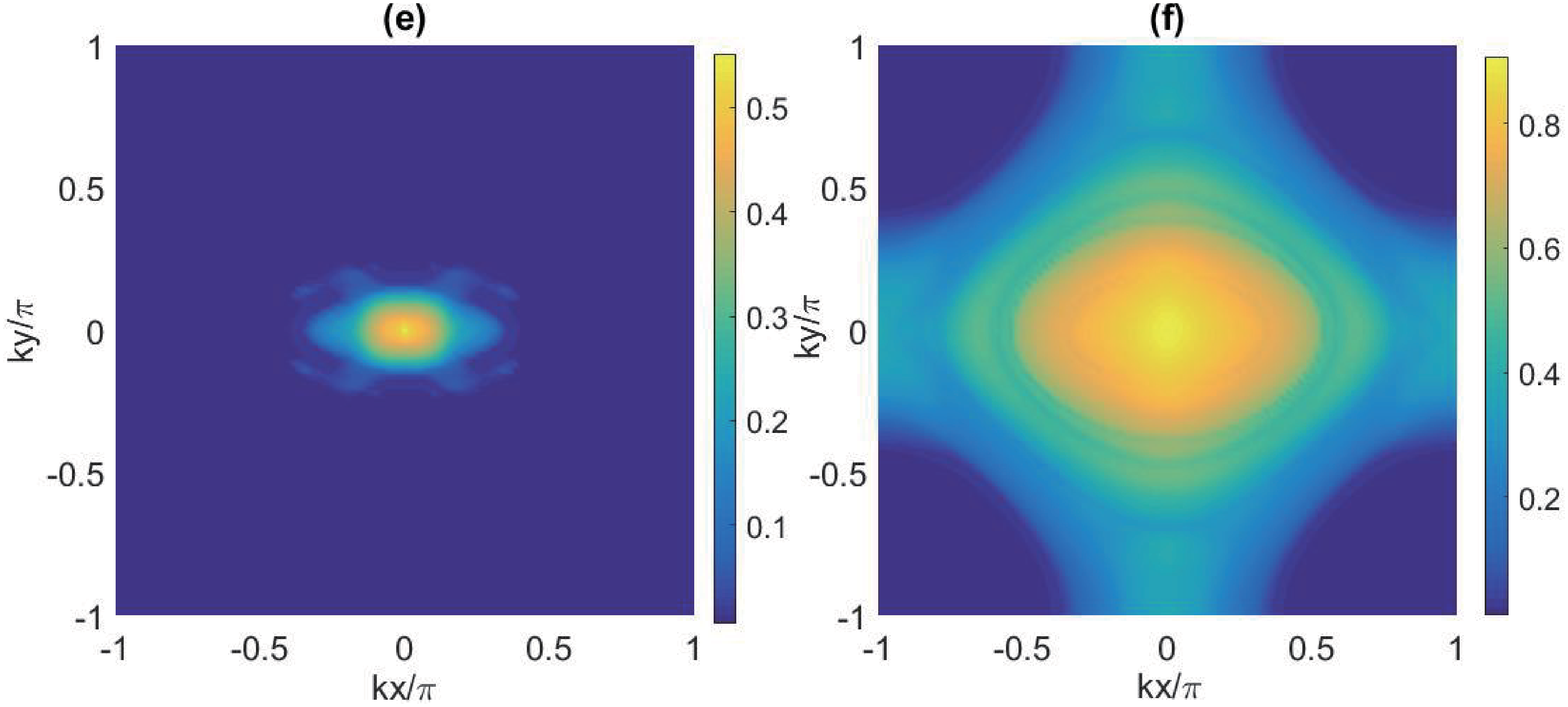}
\caption{\label{04}(Color online)
  MIPR as a function of $k_x$ and $V$ with fixed (a) $k_y=0$ and (c) $k_y=\pi$. MIPR as a function of $k_y$ and $V$ with fixed (b) $k_x=0$ and (d) $k_x=\pi$. MIPR as a function of $k_x$ and $k_y$ for fixed (e) $V=1.5$ and (f) $V=6$. Here $L=300$.}
\end{figure}
%%%%%%%%%%%%%%%%%%%%%%%%%%%%%%%%%%%%%%%%%%%%%%%%

The incommensurate potential of the $z$ direction can give rise to the extended-localization transition in this direction. To characterize it, we introduce the inverse participation ratio (IPR) \cite{Thouless,Schreiber,Evers} of the system at fixed $k_x$ and $k_y$. For a normalized wave function $\psi_n$, the IPR is defined as
$\text{IPR} = \sum_j(|\psi_{n,j,A}|^2+|\psi_{n,j,B}|^2)^2$. The IPR approaches to zero or a finite value of $O(1)$ in the thermodynamical limit corresponding to an extended or localized state. For the filled lower band, we can further define the mean IPR \cite{Cai,Hui} as
\begin{equation}
\text{MIPR}=\frac{1}{L}\sum^{L}_{n=1}\sum_j(|\psi_{n,j,A}|^2+|\psi_{n,j,B}|^2)^2,
\end{equation}
which is the average of the IPRs over all the eigenstates of the lower band.
Our results are summarized in Fig.~\ref{04}.

%%%%%%%%%%%%%%%%%%%%%%%%%%%%%%%%%%%%%%%%%%%%%%%
\begin{figure*}
\includegraphics[height=80mm,width=180mm]{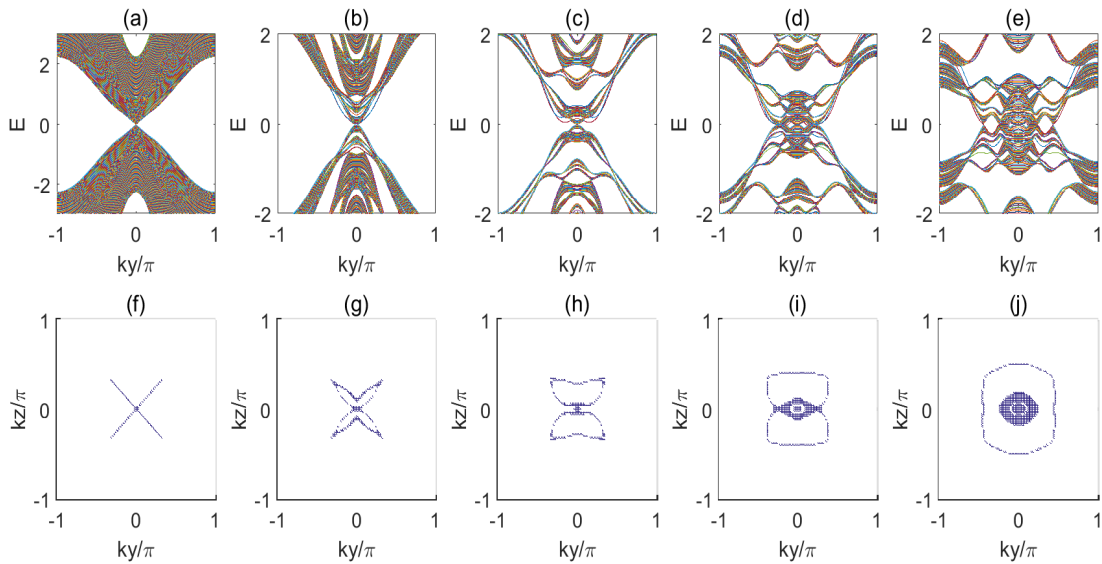}
\caption{\label{05}(Color online)
  The energy spectra as a function of $k_y$ under OBC along $x$ direction for fixed $k_z =0$ with the incommensurate potential strength (a) $V=0$, (b) $V=1$, (c) $V=2$, (d) $V=3$ and (e) $V=4$. The corresponding cross section of energy dispersion $E(k_x,k_y)$ at $E=0\pm 0.03$ for this system with (f) $V=0$, (g) $V=1$, (h) $V=2$, (i) $V=3$ and (j) $V=4$. Here $L=300$.}
\end{figure*}
%%%%%%%%%%%%%%%%%%%%%%%%%%%%%%%%%%%%%%%%%%%%%%%%

In Fig.~\ref{04}(a) and (c), we present the MIPR as a function of $k_x$ and $V$ with fixed $k_y=0$ and $k_y=\pi$ respectively. In Fig.~\ref{04}(b) and (d), we present the MIPR as a function of $k_y$ and $V$ by fixing $k_x=0$ and $k_x=\pi$ respectively. From these figures, we see that there exists an extended-localization transition when increasing the incommensurate potential strength. The states around $(k_x,k_y)=(\pm \pi,\pm \pi)$ are harder to be localized than those at other $k_x$ and $k_y$, while localization for the  states around $(k_x,k_y)=(0, 0)$ are much easier. In Fig.~\ref{04}(e), we  show the MIPR as a function of $k_x$ and $k_y$ with fixed $V=1.5$. It is clear that all eigenstates become localized
in the approximate ellipse region with the long axis $k_x \in [-0.15\pi, 0.15\pi]$ and the short axis $k_y \in [-0.1\pi, 0.1\pi]$ (see the Appendix C), which is bigger than the size of the fermi surface. In fact, the system can be viewed as a two-dimensional metal for $V\approx 1.5\sim1.8$.

In the same way, we can analyze the metallic phase for $V >4.6$ . As an example, we show the MIPR as a function of $k_x$ and $k_y$ for fixed $V=6$ in Fig.~\ref{04}(f). As Compared to Fig.~\ref{02}(h), the localized region completely contains the band cross section. Thus the system is a two-dimensional metal in this region.

Similar results can be obtained for systems with OBC in the $y$ direction and PBC in the $x$ and $z$ directions. However if we take OBC in the $x$ direction and PBC in the $y$ and $z$ directions,  things would be different. In this case, $k_y$ and $k_z$ are good quantum numbers. A crucial difference is that there exist no edge states on the surface BZ spanned by $(k_y,k_z) $ at $V=0$. The energy spectra as a function of $k_y$ for fixed $k_z=0$ with different incommensurate potential strengths are shown in Fig.~\ref{05}(a) $V=0$, (b) $V=1$, (c) $V=2$, (d) $V=3$ and (e) $V=4$. Corresponding, we also demonstrate the cross sections of energy dispersion $E(k_x,k_y)$ at $E=0\pm 0.03$  in Figs.~\ref{05}(f)-(j).

In the absence of incommensurate potential, $V=0$, the cross section of energy dispersion $E(k_x,k_y)$ at $E=0\pm 0.03$ is two lines with no surface states as shown in Fig.~\ref{05}(a) and (f). By increasing $V$, one line is broken into two with unchanged end points as shown in Fig.~\ref{05}(b) and (g). Further increasing $V$, e.g., $V=2$ as shown in Fig.~\ref{05}(c) and (h), The Fermi surface becomes two loops with a touching point at $k_x=0,k_y=0$. Keep increasing the incommensurate potential strength, e.g., $V=3$ as shown in Fig.~\ref{05}(d) and (i), this system enters into a metallic phase with the emergence of middle Fermi surface. Further increasing $V$, e.g., $V=4$ as shown in Fig.~\ref{05}(e) and (j), the two loops would merge into one loop, separated from the middle surface.

%%%%%%%%%%%%%%%%%%%%%%%%%%%%
\section{Summary}
\label{conclusion}
%%%%%%%%%%%%%%%%%%%%%%%%%%%%%%%
In summary, we have investigated the effect of  one-dimensional incommensurate potential on the nodal-link semimetals. When the potential is added along
$z$ ($y$) direction, there exist surface states on the 2D surface BZ. The system undergoes several phase transition by increasing the strength of incommensurate potential: first from a nodal-link semimetallic phase to a metallic phase, then to a loop semimetal phase and a metallic phase again. When the potential is added along $x$ direction and no surface states exist on the 2D surface BZ, the system will form two loops from the initally two Fermi lines and then the two loops merge into one closed loop.  In this process, the middle closed point would not  open and the system enter directly into a metallic phase.
%%%%%%%%%%%%%%%%%
\begin{acknowledgments}
The work is supported by NSFC under Grants No. 11425419, the National Key Research and Development Program of China (2016YFA0300600 and 2016YFA0302104) and the Strategic Priority Research Program (B) of the Chinese Academy of Sciences  (No. XDB07020000).
\end{acknowledgments}
%%%%%%%%%%%%%%%%%%%%%%%
\appendix
%%%%%%%%%%%%%%%%%%%%%%%%%%%%
\section{}
%%%%%%%%%%%%%%%%%%%%%%%%%%%%%%%
In this appendix, we give some details of derivation of eigenvalue equations from Eqs. (\ref{ham-1}), (\ref{ham-2}) and (\ref{ham-3}). Since $k_x$ and $k_y$ are good quantum numbers, we take the Fourier transform: $a^{\gamma}_{k_x,k_y,l}=\frac{1}{\sqrt{L_z}}\sum_{k_z}e^{i(k_zla)}a^{\gamma}_{k_x,k_y,k_z}$, where $\gamma=A, B$ and $a$ is the lattice constant. After the Fourier transform, we rewrite the Eq. (\ref{ham-3}) as:
\[
H=\sum_{k_x,k_y} H(k_x,k_y),
\]
with
\begin{widetext}
\bea
 H(k_x,k_y) &=& \sum_{l}\left[ 2\sin k_y(m_0-\cos k_x-\cos k_y) a^{A\dagger}_{k_x,k_y,l}a^{B}_{k_x,k_y,l} \right. \nonumber \\
            & &   + (m_0-\cos k_x-\cos k_y)a^{A\dagger}_{k_x,k_y,l+1}a^{A}_{k_x,k_y,l}-(m_0-\cos k_x-\cos k_y)a^{B\dagger}_{k_x,k_y,l+1}a^{B}_{k_x,k_y,l}  \nonumber\\
            & &   + (-\sin k_y-i\sin k_x)a^{A\dagger}_{k_x,k_y,l+1}a^{B}_{k_x,k_y,l}+(-\sin k_y+i\sin k_x)a^{A\dagger}_{k_x,k_y,l-1}a^{B}_{k_x,k_y,l}+H.c. \nonumber\\
            & &   + (\sin^2k_x-\sin^2k_y-(m_0-\cos k_x-\cos k_y)^2-1)(n^{A}_{k_x,k_y,l} - n^{B}_{k_x,k_y,l}) \nonumber\\
 & & \left.  + V \cos(2\pi\alpha l)(n^{A}_{k_x,k_y,l} + n^{B}_{k_x,k_y,l}) + H.c. \right] \nonumber
\eea
\end{widetext}
where $a^{A}_{k_x,k_y,l}$ and $a^{B}_{k_x,k_y,l}$ ($a^{A\dagger}_{k_x,k_y,l}$ and $a^{B\dagger}_{k_x,k_y,l}$) are the annihilation (creation) operators on the sublattice A and B of the $l-th$ layer along the z direction for fixed $k_x$ and $k_y$, $n^{A}_{k_x,k_y,l}=a^{A\dagger}_{k_x,k_y,l}a^{A}_{k_x,k_y,l}$ and $n^{B}_{k_x,k_y,l}=a^{B\dagger}_{k_x,k_y,l}a^{B}_{k_x,k_y,l}$.
The $n-th$ eigenstate $|\Psi_n\rangle$ of $H(k_x,k_y)$ can be represented as
\begin{equation}
|\Psi_n\rangle=\sum_{l}(\psi_{n,l,A}a^{A\dagger}_{k_x,k_y,l}+\psi_{n,l,B}a^{B\dagger}_{k_x,k_y,l})|0\rangle.
\end{equation}
then we use $H(k_x,k_y) |\Psi_n\rangle=E_n(k_x,k_y)|\Psi_n\rangle$ and obtain
\begin{widetext}
\begin{equation}
\begin{split}
&E_n\psi_{n,j,A}=(m_0-\cos k_x-\cos k_y)(\psi_{n,j-1,A}+\psi_{n,j+1,A})+(i\sin k_x-\sin k_y)\psi_{n,j+1,B}-(i\sin k_x+\sin k_y)\psi_{n,j-1,B}\\
&+(\sin^2 k_x+\sin^2 k_y-(m_0-\cos k_x-\cos k_y)^2-1+V \cos(2\pi\alpha j))\psi_{n,j,A}+2\sin k_y(m_0-\cos k_x-\cos k_y)\psi_{n,j,B},\\
&E_n\psi_{n,j,B}=-(m_0-\cos k_x-\cos k_y)(\psi_{n,j-1,B}+\psi_{n,j+1,B})+(i\sin k_x-\sin k_y)\psi_{n,j+1,A}-(i\sin k_x+\sin k_y)\psi_{n,j-1,A}\\
&+(-\sin^2 k_x-\sin^2 k_y+(m_0-\cos k_x-\cos k_y)^2+1+V \cos(2\pi\alpha j))\psi_{n,j,B}+2\sin k_y(m_0-\cos k_x-\cos k_y)\psi_{n,j,A},
\label{appen1}
\end{split}
\end{equation}
\end{widetext}
which $j$ is the $j-th$ layer along the $z$ direction of the system and $E_{n,k_x,k_y}$ is the $n$-th eigenvalue of the Hamiltonian at $(k_x,k_y)$. In a similar way, we can obtain the eigenvalue equation for the case of taking OBC along $x$ or $y$ direction.

After introducing a vector $\Psi= (\psi_{n,1,A},\psi_{n,1,B},\psi_{n,2,A},\psi_{n,2,B},\cdots, \psi_{n,L,A},\psi_{n,L,B})$, Eqs. (\ref{appen1}) can be written as a $2L\times 2L$ matrix:
\begin{equation*}
 H(k_x,k_y) =
\begin{pmatrix}
 A_1&B&0&\cdots& & &0 &0\\
B^{\dagger}&A_2&B&0&\cdots& & &0\\
0&B^\dagger&A_3&B&0&\cdots& &0\\
0&0&B^\dagger&A_4&B&0&\cdots&0\\
\vdots&\ddots&\ddots&\ddots&\ddots&\ddots&\ddots&\vdots\\
0& &\cdots &0 &B^\dagger &A_{L-2} &B &0\\
0& & &\cdots &0 &B^\dagger &A_{L-1} &B\\
0& & & &\cdots &0 &B^\dagger &A_{L}
\end{pmatrix},
%\eea
\end{equation*}
\\
where
%\begin{widetext}
\bea
 A_j=
 \begin{pmatrix}
 A_{11} & A_{12}\\
A_{21} & A_{22}
\end{pmatrix}
\notag
\eea
with $A_{11}=\sin^2 k_x+\sin^2 k_y-(m_0-\cos k_x-\cos k_y)^2-1+V \cos(2\pi\alpha j)$, $A_{12}=2\sin k_y(m_0-\cos k_x-\cos k_y)$, $A_{21}=2\sin k_y(m_0-\cos k_x-\cos k_y)$ and
 $A_{22}=-\sin^2 k_x-\sin^2 k_y+(m_0-\cos k_x-\cos k_y)^2+1+V \cos(2\pi\alpha j)$,
%\end{widetext}
and
\bea
 B=
\begin{pmatrix}
 m_0-\cos k_x-\cos k_y & i\sin k_x-\sin k_y\\
 i\sin k_x-\sin k_y & -m_0+\cos k_x+\cos k_y
\end{pmatrix}.
\notag
\eea
By numerically diagonalizing the matrix, we can obtain the eigenvalues and eigenstates of the Hamiltonian (\ref{ham-3}).

%%%%%%%%%%%%%%%%%%%%%%%%%%%%%%%%%%%%%%%%%%%%%%%
\begin{figure}
\includegraphics[height=75mm,width=85mm]{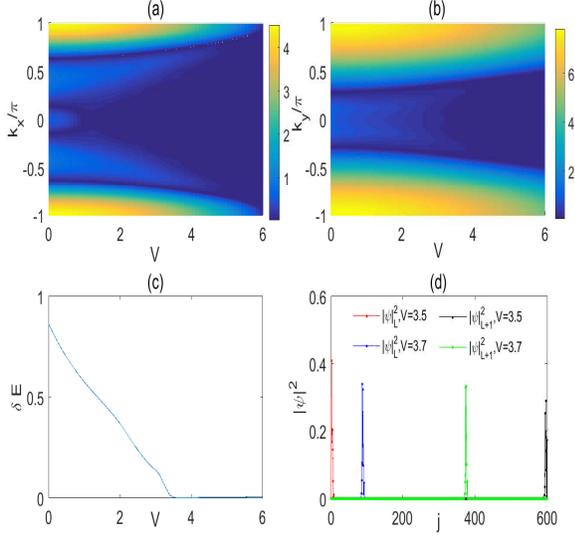}
\caption{\label{07}(Color online)
 (a) $\delta E$ as a function of $k_x$ and $V$ with $k_y=0$, (b) $\delta E$ as a function of $k_y$ and $V$ with $k_x=0.4\pi$, (c) $\delta E$ versus $V$ with fixed $k_y=0$ and $k_x=0.4\pi$, (d) The distribution of the $L-th$ and $(L+1)-th$ eigenstates of this system with $V=3.5$ and $V=3.7$, here we fixed $k_y=0$ and $k_x=0.4\pi$. Here $L=300$.}
\end{figure}
%%%%%%%%%%%%%%%%%%%%%%%%%%%%%%%%%%%%%%%%%%%%%%%%

%%%%%%%%%%%%%%%%%%%%%%%%%%%%
\section{}
%%%%%%%%%%%%%%%%%%%%%%%%%%%%%%%
From Fig.~\ref{02}(a) and (b), we can see that the surface states are absorbed into the lower band with increasing the strength of the incommensurate potential. To explain this in detail, we define a energy gap $\delta E_{k_x,k_y} = E_{L+1,k_x,k_y}- E_{L-1,k_x,k_y}$, where $E_{L+1,k_x,k_y}$ and $E_{L-1,k_x,k_y}$ are the $(L+1)-th$ and $(L-1)-th$ eigenvalues with $(k_x,k_y)$ and the eigenvalues are in ascending order. For fixed $k_x$ and $k_y$, when the surface states are absorbed into the lower band or the gap between the lower band and the upper band near $E = 0$ closes, $\delta E_{k_x,k_y}$ will equal to zero.
Fig.~\ref{07}(a) shows $\delta E$ as a function of $k_x$ and $V$ with $k_y=0$, one can see that $\delta E$ changes from a non-zero value to zero near
$k_x=\pm 0.4\pi$ with increasing the incommensurate potential strength $V$. As shown in Fig.~\ref{01}(c) and (d), there exists surface states near $k_x=\pm 0.4\pi$ and the the gap between the lower band and the upper band near $E = 0$ doesn't close when $V\approx 4$ near $k_x=\pm 0.4\pi$ as shown in Fig.~\ref{01s} and Fig.~\ref{02}(c), so $\delta E$ approaching to zero indicates that the surface states are absorbed into the lower band. Fig.~\ref{07}(b) shows $\delta E$ as a function of $k_y$ and $V$ with $k_x=0.4\pi$. Similarly, $\delta E$ changes from a non-zero value to zero near $k_y=0$ when increasing $V$, which indicates that the surface states are absorbed into the lower band.

To be specific, we fix $k_y=0$ and $k_x=0.4\pi$. Fig.~\ref{07}(c) displays $\delta E$ as a function of $V$, one can see that $\delta E$ changes from a non-zero value to zero at about $V=3.6$. In fig.~\ref{07}(d), we display the distribution of the $L-th$ and $(L+1)-th$ eigenstates with $V=3.5$ and $V=3.7$, one can see that the distributions of the two states are located at the boundaries of $z$ direction when $V=3.5$. On the contrary, distributions
of the two states locate inside of the bulk of $z$ direction when $V=3.7$, which indicates the disappearance of the boundary states for $k_y=0$ and $k_x=0.4\pi$. For other $k_x$ and $k_y$, we can make a similar analysis and obtain the same conclusion that the surface states vanish precisely because they are absorbed into the lower band with increasing the strength of the incommensurate potential.

%%%%%%%%%%%%%%%%%%%%%%%%%%%%%%%%%%%%%%%%%%%%%%%
\begin{figure}
\includegraphics[height=80mm,width=85mm]{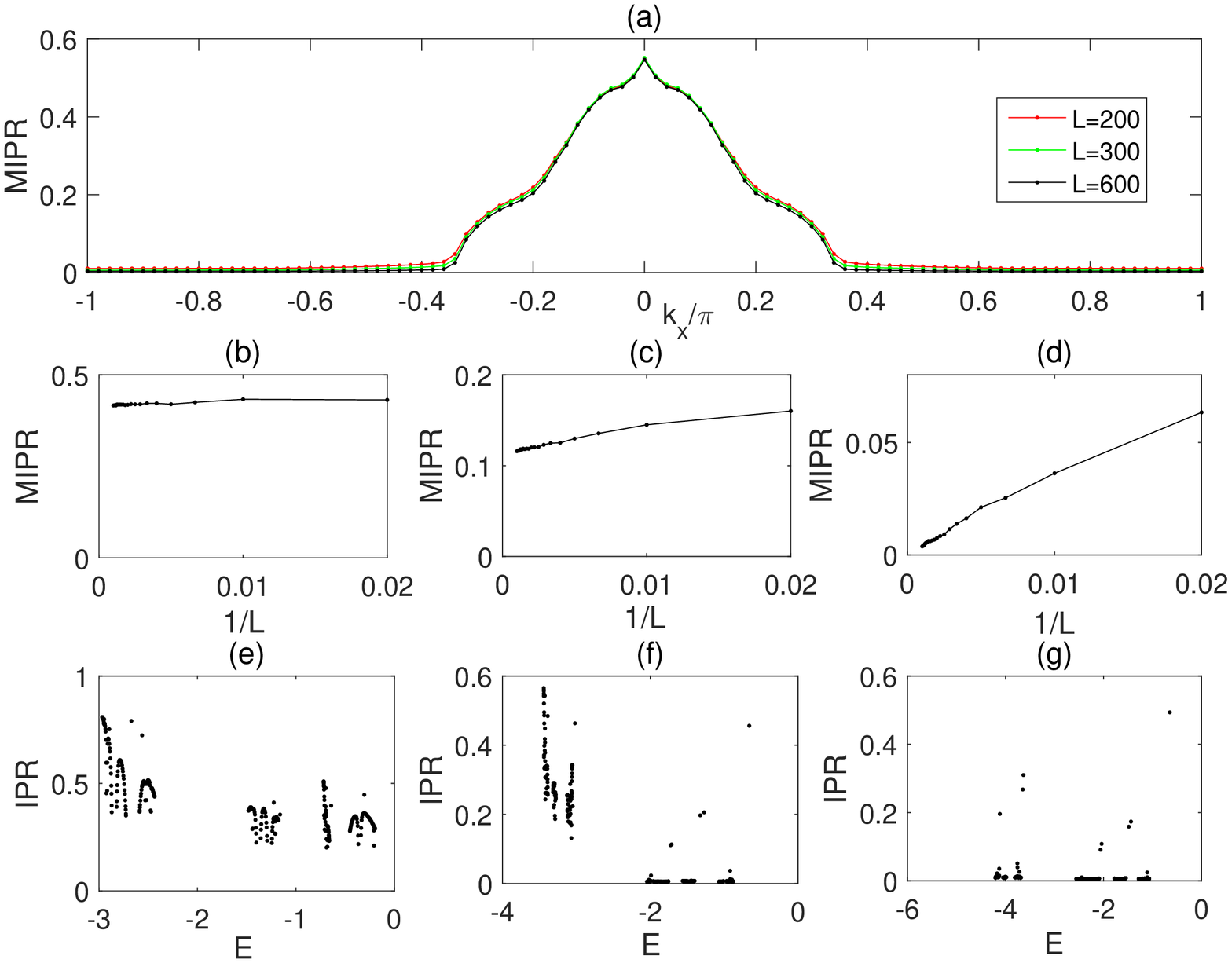}
\caption{\label{06}(Color online)
 (a) MIPR as a function of $k_x$ for different sizes $L$. MIPR versus $1/L$ with fixed (b) $k_x=0.1\pi$, (c) $k_x=0.3\pi$ and (d) $k_x=0.4\pi$. IPR versus eigenvalue $E$ for fixed (e) $k_x=0.1\pi$, (f) $k_x=0.3\pi$ and (g) $k_x=0.4\pi$. Other parameters are $V=1.5$ and $k_y=0$.}
\end{figure}
%%%%%%%%%%%%%%%%%%%%%%%%%%%%%%%%%%%%%%%%%%%%%%%%
%%%%%%%%%%%%%%%%%%%%%%%%%%%%
\section{}
%%%%%%%%%%%%%%%%%%%%%%%%%%%%%%%
In this appendix, we use $V=1.5$ as an example to discuss some details about the IPR. In Fig.~\ref{04}(e), we showed the MIPR as a function of $k_x$ and $k_y$ with fixed $V=1.5$. Now we fix $k_y=0$ and present the MIPR as a function of $k_x$ for $L=200,300,600$ in Fig.~\ref{06}(a). We see that the MIPR is close to zero and it becomes smaller with the system size increased when $|k_x|$ is bigger than $0.4\pi$ but the MIPR is a finite value near $0.5$ and it remains almost unchanged with increasing $L$ when $|k_x|$ is small. To further explain this point, we present MIPR as a function of $1/L$ with fixed $k_x=0.1\pi, 0.3\pi$ and $0.4\pi$ in Fig.~\ref{06}(b), (c) and (d) respectively. We see that MIPR is almost unchanged with increasing $L$ for $k_x=0.1\pi$. It slightly change when $L$ is increased for $k_x=0.3\pi$ but it doesn't become zero when $L$ approaches infinity. It diminishes with $L$ increased for $k_x=0.4\pi$ and it will tend to zero with $L\rightarrow \infty$. Moreover, we show the corresponding IPR as a function of eigenvalues $E$ in Fig.~\ref{06} (e), (f) and (g) respectively. We see that all IPRs are non-zero values when $k_x=0.1\pi$, which means that all eigenstates are localized. Some IPRs are non-zero and some approximate to zero for $k_x=0.3\pi$, which indicates the existence of a mobility edge. One can notice that there exist some IPRs that are non-zero but their neighbor ones are close to zero. These states actually correspond to edge states in the energy gaps of this system \cite{Wangy}. For $k_x=0.4\pi$, all IPRs approximate to zero except the IPRs that some edge states correspond to. According to the similar analysis, we can obtain that all eigenstates are localized in the approximate ellipse region with the long axis $k_x \in [-0.15\pi, 0.15\pi]$ and the short axis $k_y \in [-0.1\pi, 0.1\pi]$.

%\clearpage
%%%%%%%%%%%%%%%%%%%%%%%%%%%%%%%%%%%%%%%%%%%%%


\begin{thebibliography}{36}
\bibitem{Nagaosa} B.-J. Yang and N. Nagaosa, Nat. Commun. {\bf 5}, 4898 (2014).
\bibitem{Fang} C. Fang, Y. Chen, H.-Y. Kee, and L. Fu, Phys. Rev. B {\bf 92}, 081201 (2015).
\bibitem{Kane} M. Z Hassan and C. L. Kane, Rev. Mod. Phys. {\bf 82}, 3045 (2010).
\bibitem{Zhang} X. L. Qi and S.-C. Zhang, Rev. Mod. Phys. {\bf 83}, 1057 (2011).
\bibitem{Chiu} C. K. Chiu, J. C. Y. Teo, A. P. Schnyder, and S. Ryu,
Rev. Mod. Phys. {\bf 88}, 035005 (2016).
\bibitem{class1} A. P. Schnyder, S. Ryu, A. Furusaki, and A. W. W. Ludwig, Phys. Rev. B {\bf 78}, 195125 (2008).
\bibitem{class2} S. Ryu, A. P. Schnyder, A. Furusaki, and A. W. W. Ludwig, New J. Phys. {\bf 12}, 065010 (2010).
%\bibitem{class1} A. P. Schnyder, S. Ryu, A. Furusaki, and A. W. W. Ludwig, \href{http://link.aps.org/doi/10.1103/PhysRevB.78.195125}Phys. Rev. B {\bf 78}, 195125 (2008).
%\bibitem{class2} S. Ryu, A. P. Schnyder, A. Furusaki, and A. W. W. Ludwig, \href{http://dx.doi.org/10.1088/1367-2630/12/6/065010}New J. Phys. {\bf 12}, 065010 (2010).
\bibitem{Murakami} S. Murakami,  New J. Phys. {\bf 9}, 356 (2007).
\bibitem{Wan} X. Wan, A. M. Turner, A. Vishwanath, and S. Y. Savrasov,
Phys. Rev. B {\bf 83}, 205101 (2011).
\bibitem{Burkov} A. A. Burkov and L. Balents, Phys. Rev. Lett. {\bf 107}, 127205 (2011).
\bibitem{Yang} K.-Y. Yang, Y.-M. Lu, and Y. Ran, Phys. Rev. B {\bf 84}, 075129 (2011).
\bibitem{Xu} G. Xu, H. Weng, Z. Wang, X. Dai, and Z. Fang, Phys.
Rev. Lett. {\bf 107}, 186806 (2011).
\bibitem{Halasz} G. B. Halasz and L. Balents, Phys. Rev. B {\bf 85}, 035103 (2012).
\bibitem{Young} S. M. Young, S. Zaheer, J. C. Y. Teo, C. L. Kane, E. J.
Mele, and A. M. Rappe, Phys. Rev. Lett. {\bf 108}, 140405 (2012).
\bibitem{Huang} S.-M. Huang, S.-Y. Xu, I. Belopolski, C.-C. Lee, G.
Chang, B. Wang, N. Alidoust, G. Bian, M. Neupane,
C. Zhang, S. Jia, A. Bansil, H. Lin, and M. Z. Hasan,
Nat. Commun. {\bf 6}, 7373 (2015).
\bibitem{Lv} B. Q. Lv, N. Xu, H. M. Weng, J. Z. Ma, P. Richard, X.
C. Huang, L. X. Zhao, G. F. Chen, C. E. Matt, F. Bisti,
V. N. Strocov, J. Mesot, Z. Fang, X. Dai, T. Qian, M.
Shiand, and H. Ding, Nat. Phys. {\bf 11}, 724 (2015).
\bibitem{Hasan} S.-Y. Xu, I. Belopolski, N. Alidoust, M. Neupane, G.
Bian, C. Zhang, R. Sankar, G. Chang, Z. Yuan, C.-C.
Lee, S.-M. Huang, H. Zheng, J. Ma, D. S. Sanchez, B.
Wang, A. Bansil, F. Chou, P. P. Shibayev, H. Lin, S. Jia,
and M. Z. Hasan, Science {\bf 349}, 613 (2015).
\bibitem{Hasan2} S.-Y. Xu, N. Alidoust, I. Belopolski, Z. Yuan, G. Bian,
T.-R. Chang, H. Zheng, V. N. Strocov, D. S. Sanchez, G. Chang, C. Zhang, D. Mou, Y. Wu, L. Huang, C.-C.
Lee, S.-M. Huang, B.-K. Wang, A. Bansil, H.-T. Jeng, T.
Neupert, A. Kaminski, H. Lin, S. Jia, and M. Z. Hasan,
Nat. Phys. {\bf 11}, 748 (2015).
\bibitem{Yang2} L. X. Yang, Z. K. Liu, Y. Sun, H. Peng, H. F. Yang,
T. Zhang, B. Zhou, Y. Zhang, Y. F. Guo, M. Rahn, D.
Prabhakaran, Z. Hussain, S.-K. Mo, C. Felser, B. Yan,
and Y. L. Chen, Nat. Phys. {\bf 11}, 728 (2015).
\bibitem{Burkov2} A. A. Burkov, M. D. Hook, and L. Balents, Phys. Rev. B {\bf 84}, 235126 (2011).
\bibitem{Fang2} C. Fang, H. Weng, X. Dai, and Z. Fang, Chin. Phys. B {\bf 25}, 117106 (2016).
\bibitem{Chan2016} Y.-H. Chan, C.-K. Chiu, M. Y. Chou, and A. P. Schnyder, Phys. Rev. B 93, 205132 (2016).

\bibitem{Hou} W. Chen, H.-Z. Lu, and J.-M. Hou, Phys. Rev. B {\bf 96}, 041102(R) (2017).
\bibitem{Yan} Z. Yan, R. Bi, H. Shen, L. Lu, S.-C. Zhang, and Z. Wang, Phys. Rev. B {\bf 96}, 041103(R) (2017).
\bibitem{Chang} P.-Y. Chang, and C.-H. Yee, arXiv: 1704.01948.
\bibitem{Ezawa} M. Ezawa, Phys. Rev. B {\bf 96}, 041202 (2017).
\bibitem{DuanLM17} X.-X. Yuan, L. He, S.-T. Wang, D.-L. Deng, F. Wang, W.-Q. Lian, X. Wang, C.-H. Zhang, H.-L. Zhang, X.-Y. Chang, L.-M. Duan,  arXiv:1705.00781.
\bibitem{Herbut} K. Kobayashi, T. Ohtsuki, K.-I. Imura, and I. F. Herbut, Phys. Rev.
Lett. {\bf 112}, 016402 (2014).
\bibitem{Sondhi} R. Nandkishore, D. A. Huse, and S. L. Sondhi, Phys. Rev. B {\bf 89}, 245110 (2014).
\bibitem{Sarma1} B. Roy and S. Das Sarma, Phys. Rev. B 90, 241112 (2014).
\bibitem{Sarma2} J. H. Pixley, P. Goswami, and S. D. Sarma, Phys. Rev. Lett. {\bf 115},
076601 (2015).
\bibitem{Huang2} Z. Huang, T. Das, A. V. Balatsky, and D. P. Arovas, Phys. Rev. B {\bf 87}, 155123 (2013).
\bibitem{Ominato} Y. Ominato and M. Koshino, Phys. Rev. B {\bf 89}, 054202 (2014).
\bibitem{Brouwer} B. Sbierski, G. Pohl, E. J. Bergholtz, and P. W. Brouwer, Phys. Rev. Lett. {\bf 113}, 026602 (2014).
\bibitem{Ryu} R. R. Biswas and S. Ryu, Phys. Rev. B {\bf 89}, 014205 (2014).
\bibitem{Xie} C.-Z. Chen, J. Song, H. Jiang, Q.-F. Sun, Z. Wang, and X. C. Xie,
Phys. Rev. Lett. {\bf 115}, 246603 (2015);
Y. Wu, H. Liu, H. Jiang, and X. C. Xie,
Phys. Rev. B {\bf 96}, 024201 (2017).
\bibitem{Hughes} H. Shapourian and T. L. Hughes, Phys. Rev. B {\bf 93}, 075108 (2016).
\bibitem{Roy} S. Bera, J. D. Sau, and B. Roy, Phys. Rev. B {\bf 93}, 201302 (2016).
\bibitem{Gurarie} L. Radzihovsky, L. Radzihovsky, and V. Gurarie,  Phys. Rev. Lett.  {\bf 114}, 166601 (2015);
S.V. Syzranov and L. Radzihovsky, arXiv:1609.05694.
\bibitem{Wang} Y. Wang and S. Chen, Phys. Rev. A {\bf 95}, 053634 (2017).
\bibitem{Pixley} J. H. Pixley, J. H. Wilson, D. A. Huse, and S. Gopalakrishnan, Phys. Rev. Lett. {\bf 120}, 207604 (2018).
\bibitem{Roati} G. Roati, C. D'Errico, L. Fallani, M. Fattori, C. Fort, M. Zaccanti, G. Modugno, M. Modugno, and M. Inguscio, Nature {\bf 453}, 895 (2008).
\bibitem{Sokoloff} J. Sokoloff, Phys. Rep. {\bf 126}, 189 (1985).
\bibitem{Azbel} M. Y. Azbel, Phys. Rev. Lett. 43, 1954 (1979).
\bibitem{AA} S. Aubry and G. Andr\'{e}, Ann. Isr. Phys. Soc. {\bf 3}, 133 (1980).
\bibitem{Takane} Y. Takane, J. Phys. Soc. Jpn. {\bf 85}, 124711 (2016).

\bibitem{Zhao} Y. X. Zhao and Y. Lu, Phys. Rev. Lett. {\bf 118}, 056401 (2017).
\bibitem{Moore} J. E Moore, Y. Ran, and X.-G. Wen, Phys. Rev. Lett. {\bf 101}, 186805 (2008).
\bibitem{Deng1} D.-L. Deng, S.-T. Wang, C. Shen, and L.-M. Duan, Phys. Rev. B {\bf 88}, 201105 (2013).
\bibitem{Deng2} D.-L. Deng, S.-T. Wang, K. Sun, and L.-M. Duan, arXiv:1612.01518.
\bibitem{Deng3} D.-L. Deng, S.-T. Wang, and L.-M. Duan, Phys. Rev. B {\bf 89}, 075126 (2014).
\bibitem{Kennedy} R. Kennedy, Phys. Rev. B {\bf 94}, 035137 (2016).
\bibitem{Liu} C. Liu, F. Vafa, and C. Xu, Phys. Rev. B {\bf 95}, 161116(R) (2017).
\bibitem{Fehske} A. Wei\ss e, G. Wellein, A. Alvermann, and H. Fehske, Rev. Mod. Phys. {\bf 78}, 275 (2006).
\bibitem{Kohmoto} M. Kohmoto and D. Tobe, Phys. Rev. B {\bf 77}, 134204 (2008);
 M. Kohmoto, Phys. Rev. Lett. {\bf 26}, 1198 (1983); C. Tang and M. Kohmoto, Phys. Rev. B {\bf 34}, 2041 (1986).
\bibitem{Thouless} D. J. Thouless, Phys. Rep. {\bf 13}, 93 (1974).
\bibitem{Schreiber} M. Schreiber, J. Phys. C {\bf 18}, 2493 (1985); Y. Hashimoto,
K. Niizeki, and Y. Okabe, J. Phys. A {\bf 25}, 5211 (1992).
\bibitem{Evers} F. Evers and A. D. Mirlin, Phys. Rev. Lett. {\bf 84}, 3690 (2000).
\bibitem{Cai} X. Cai, L.-J. Lang, S. Chen, and Y. Wang,  Phys. Rev. Lett. {\bf 110}, 176403 (2013).
\bibitem{Hui} J. Wang, X.-J. Liu, G. Xianlong and H. Hu, Phys. Rev. B {\bf 93}, 104504 (2016).
\bibitem{Wangy} Y. Wang, G. Xianlong and S. Chen, Eur. Phys. J. B {\bf 89}, 254 (2016).
\end{thebibliography}
\end{document}